\documentclass[11pt]{article}

\usepackage{amsmath}
\usepackage{amsfonts}
\usepackage{amssymb}\usepackage{amsmath}
\usepackage{amsfonts}
\usepackage{amssymb}

\begin{document}
\pagestyle{headings}
\renewcommand{\thefootnote}{\alph{footnote}}

\title{Are strings the aether of our time?}
\author{Glenn Eric Johnson\\Oak Hill, VA.\\E-mail: glenn.e.johnson@gmail.com}
\maketitle

{\bf Abstract:} Descriptions of relativistic quantum physics that derive from quantizations of classical physics require additional technical properties and these technical conjectures exclude interaction in example realizations. In this essay, uniquely quantum mechanical examples are discussed to illustrate realizations made available by displacing classical concepts from relativistic quantum physics.


\section{Introduction}

The title is intended to provoke questions about the use of classical concepts, the example is strings, in descriptions of relativistic quantum physics. Why are distinguishable, classical geometric objects considered to describe relativistic quantum dynamics? As the more general description of nature, quantum mechanics should stand on its own without a foundation in classical concepts. While it is our experience that classical limits of quantum mechanics are usefully interpreted using the geometric concepts of point or string in spacetime, this interpretation applies to a limited set of approximations to quantum physics. For example, classical gravity predicts spacetime singularities while one perspective of quantum physics is that the interpretation of states as distinguishable objects falling into a spacetime singularity fails at small distances. The description of states as elements of a Hilbert space remains when the approximation and the geometric interpretation fails. The classical approximation develops a singularity while the appropriate description of physical states as Hilbert space elements continues a unitary evolution. Indeed, in the rigged Hilbert spaces of interest in relativistic quantum physics, the arguments of functions that label states are variables of integration in representations of generalized functions\footnote{Chapter II, section 4 in [\ref{gel2}]}. A spacetime geometry is the result of interpretation of particular states that are perceived to evolve along trajectories on manifolds according to classical models. The scalar product provides the geometry native to the Hilbert space of states. The intended analogy with aether is to emphasize that effective methods are not necessarily elaborations of prevailing methods. At one time an aether that exhibited electromagnetic waves and had properties consistent with the observed independence of the speed of light from an observer's velocity was of interest. This inquiry was not insightful, and physics abandoned it to describe electromagnetism without considering electromagnetic waves as disturbances in a medium. Similarly, limiting descriptions of relativistic quantum physics to quantizations of classical models may unnaturally constrain development. Indeed, quantum mechanics developed when observation contradicted the classical concept of distinguishable objects having trajectories.


Of course, difficulties in the canonical quantization of relativistic field theory were recognized from the start [\ref{wightman-hilbert}] and mitigations such as S-matrix and string theories [\ref{witten}] have been suggested. Feynman series [\ref{weinberg}] provide a phenomenologically successful development for relativistic quantum physics and the analysis of diagrams appeals to classical intuition but whether the development realizes quantum mechanics remains undemonstrated. In contrast, there are Hilbert space realizations of relativistic quantum physics that exhibit interaction in physical spacetimes that are not derived from, and indeed are excluded by, a characterization as canonical quantizations [\ref{intro}]. These constructions are uniquely quantum mechanical, possess anticipated classical limits [\ref{limits}], include $1/r$ and Yukawa equivalent potentials [\ref{feymns}], and display substantive characteristics of nature: causality, invariance of transition amplitudes under Poincar\'{e} transformations, nonnegative energy, and a Hilbert space realization. Scattering amplitudes from the constructions have asymptotes at weak coupling and small momentum exchanges that approximate Feynman series amplitudes. But, the constructions that exhibit interaction violate prevailing technical conjecture for quantum field theories: the real fields are not Hermitian Hilbert space field operators, no states have strictly bounded spacetime support, and the interaction Hamiltonians vanish while interaction is manifest. The constructions would not be considered as realizations of quantum field theory in prevailing developments. Nevertheless, the constructions provide explicit examples of alternatives to canonical quantizations. Although abandonment of canonical quantization leaves physics without familiar characterizations for dynamics, the abandonment achieves realizations of relativistic quantum physics that exhibit familiar interactions in physical spacetime. Explicit examples enable critical review of preconceptions. For example, the constructions escape Haag's theorem in an unanticipated way: in the example of a single Lorentz scalar field, the constructions have a two-point function that can be taken to be the Pauli-Jordan positive frequency function but with a field that is not a Hermitian Hilbert space operator. As a consequence, the field is not unitarily similar to the Hermitian free field operator. The constructions abandon reliance on quantizations of classical models and perturbation of free fields. Consistency of classical limits with models of classical physics is a weaker constraint than requiring that quantum dynamics results from quantizations of classical limits.


Association of classical dynamic variables with self-adjoint Hilbert space operators is one questionable extrapolation of canonical quantization to relativistic quantum physics. This association provides a translation from classical to quantum mechanics but the extrapolation to relativity and gravity results in contradictions and requires generalizations that are not evidently motivated except to mitigate those contradictions. Association of classical dynamic variables with self-adjoint Hilbert space operators is essential to canonical quantization and canonical quantization is used despite counterexamples to the necessity of such an association. Indeed, due to Lorentz covariance, location is not directly associated with a self-adjoint operator [\ref{wigner},\ref{yngvason-lqp},\ref{johnson}]. That is, multiplications by the values of arguments of functions that label states do not correspond to self-adjoint Hilbert space operators except in nonrelativistic limits. The successes of nonrelativistic quantum mechanics and Feynman series are indications that quantum physics constrained to have appropriate classical limits does approximate nature. Nevertheless, the constructions suggest that inclusion of relativity with interaction is inconsistent with canonical quantizations, in particular, is inconsistent with a correspondence of relativistic classical fields with self-adjoint Hilbert space field operators. The constructions of relativistic quantum physics that exhibit interaction and trivial interaction Hamiltonians are inconsistent with a canonical quantization. Canonical quantizations are impressively successful in nonrelativistic quantum mechanics and Feynman series, and to many, these concepts are the essence of quantum mechanics, but the constructions demonstrate that relativistic quantum physics can be implemented without Hermitian Hilbert space field operators.


After nearly one hundred years of development\footnote{Counting from Max Born, {\em \"{U}ber Quantenmechanik}, {\bf Z. Phys.} 26, p.~379-395, 1924.}, controversy remains over basic understanding of quantum mechanics as the description of reality. Is quantum mechanics a new dynamics for classically idealized objects, or is quantum dynamics a distinct description of objects as elements of Hilbert spaces? Is quantum dynamics necessarily described by canonical quantizations of classical dynamics? Here, the derivation of quantum dynamics from classical dynamic models is designated a {\em quantization}, use of the canonical commutation or anticommutation relations in a quantization is a {\em canonical quantization}, and the essence of {\em classical} dynamics is a description of distinguishable geometric objects using Cartesian or locally Minkowski coordinates. Classical dynamics considers descriptions of trajectories of distinguishable objects to be reality. Quantum dynamics describes physical states as elements of Hilbert spaces with translation group homomorphic, unitary transformation determining the temporal evolution. Objects are distinguished by mass and quantum numbers, but objects are otherwise indistinguishable in quantum mechanics. The early development of quantum mechanics focused on canonical quantization and developed a correspondence of quantum with classical for nonrelativistic mechanics. However, when relativity is considered, the elevation of classical field theories to quantum mechanical theories incurs considerable difficulty: divergences in conjectured expansions, physically inequivalent realizations due to inapplicability of the Stone-von Neumann theorem, the `no go' result of the Haag (Hall-Wightman-Greenberg) theorem, and anomalies although the renormalizable development has impressive phenomenological successes. Nevertheless, even a nonrelativistic description of quantum mechanics suffers the Einstein-Podolsky-Rosen and Schr\"{o}dinger's cat paradoxes when classical concepts are retained. These measurement description paradoxes highlight the implausibility of maintaining classical descriptions of states of nature. Probabilistic, hidden variable, and ``we just don't understand it'' interpretations of quantum mechanics are characterized here as methods to preserve a classical interpretation of quantum mechanics. Alternatively, quantum mechanics is considered a complete theory in the Everett-Wheeler-Graham (EWG) relative state description [\ref{ewg}] and in axiomatic developments. Axiomatic developments rely only on explicit assumptions but realizations of relativistic quantum physics that exhibit interaction remain an unattained goal. It is argued here that this lack of realizations is due to remnants of strong, canonical quantization-based assumptions in the axioms. Examples demonstrate realizations of relativistic quantum physics that violate prevailing technical conjecture. That is, relaxation of technical conjecture originating in canonical quantization results in explicit realizations of relativistic quantum physics. These constructed examples suggest that a revisit to the foundations of quantum mechanics is productive. Free field theories, the archetype for the mathematics of relativistic quantum physics, are considered here to be misleading, isolated cases. Here, low dimensional spacetime models and free fields plus the related physically trivial models are excluded from consideration. Indeed, free fields are defined in physical spacetimes as Hermitian Hilbert space operators labeled by functions from sets that include functions of bounded spacetime support and this result has presumably contributed to the retention of technical conjecture inappropriate to interacting field theories. The constructions that exhibit interaction are singularly removed in a substantial sense from these free and related models. The constructions do not result from perturbations of free field operators. It is not suggested here that the properties of these constructions are necessary to relativistic quantum physics, and so the indicated contradictions are between the constructions and canonical quantization-based conjecture, but the longstanding lack of explicit realizations of interacting quantum fields in physical spacetimes suggests that the contradictions apply to more general cases.

The interactions of our common experience often correlate us with localized states so the classical limit dominates our experience and intuition but the truer representation of nature, one in which states can manifest in complementary ways, particles and waves, is quantum mechanics. In relativistic quantum physics, the states are elements of rigged Hilbert spaces and are labeled by functions. A Poincar\'{e} invariant scalar product necessitates that the states are labeled by functions dual to generalized functions. Relativistic Hilbert space field operators with expectation values that are functions of spacetime are inconsistent with Poincar\'{e} invariant likelihoods [\ref{bogo},\ref{wizimirski}]. The Feynman rules formalize ``interaction at a point'', an unnatural concept in relativistic quantum descriptions. The concept of a spacetime geometry for function arguments is imposed as an interpretation of the perceived trajectories of point-like states. These observations are classical approximations of the relativistic quantum physics. Particular selections for the state labels simplify the description of motion. The representations of generalized functions are selected to result in straight line trajectories for point-like, free particle states. This linear description of free motion is preserved with Poincar\'{e} transformations and scale changes. Deflections from linear motion indicates interaction. At small scales, or when packet spread or entanglement is appreciable, classical interpretations no longer necessarily apply and the evolution of the elements of the Hilbert space substitutes for the classical spacetime trajectory approximations.

Elements of Hilbert spaces describe nature. For nonrelativistic motion, these states can be considered as superpositions over classically idealized descriptions with the consideration that these classically idealized descriptions typically correspond to generalized functions and not elements of the Hilbert space. Ehrenfest's theorem provides an intuitive correspondence of classical trajectories with quantum dynamics in the case of nonrelativistic motion and point-like states, but once particle production is energetically possible, such a correspondence no longer applies. Together with consideration of observers' interpretations of states in terms of classical concepts, quantum mechanics is a complete description of nature with no need for external observers nor ad hoc collapse to particular, idealized states. The properties of operators realized in the Hilbert spaces of interest to physics are determined and their algebraic properties are not necessarily available to be specified in a canonical quantization. The generators of symmetries such as the generators of translations, energy and momentum, are self-adjoint operators but additional conjectured operators such as locations and fields are not necessarily self-adjoint Hilbert space operators. Indeed, the constructed real quantum fields are the conventional multiplication in the algebra of function sequences [\ref{borchers}] but in the case of the constructions this multiplication does not result in self-adjoint Hilbert space field operators [\ref{intro}].

The indistinguishable particles of quantum mechanics resolve the Gibbs paradox but this description is in violent conflict with intuition from considering distinguishable objects traveling trajectories. For dominantly-peaked packets, the state evolution of nonrelativistic objects is well approximated by the classical description while packet overlaps and entanglements are negligible. At a large scale for functions with isolated support concentrated in small areas, the quantum description appears point-like. Such states are designated here as {\em dominantly-peaked packet states}. However, at a small scale, or when entanglement is significant, or for multiple overlapping or energetic particles, the differences of the quantum and classical descriptions are not negligible. The scale for {\em small} here is set by the Compton wavelength, $\hbar/mc$, and corresponds with a nonrelativistic limit. Minimum packet states are the most classical states in the sense that the geometric mean of the packet extent in position and momentum is minimal. A classical limit applies to a multiple object state when dominantly-peaked behavior is exhibited in each spacetime argument and the peaks are jointly isolated. The peaks of packets with appropriate spreads propagate along classically described trajectories until spreads become too large or packets collide or bound states decay. A classical limit applies as an approximation and neglects rare events. Packet spread is negligible when the spread significantly exceeds the Compton wavelength and the dominant weight of the packet is sufficiently small to be spatially isolated [\ref{limits}]. Quantum mechanics is a description for the dynamics of elements of Hilbert spaces. The quantum state, an element of a Hilbert space, is the description of reality and paradoxes of measurement originate in attempts to consider quantum states as statistical descriptions of objects that are actually located at points, travel along trajectories, and as a consequence, go through one or the other of the two slits in a Young's double slit interferometer. The description of a particle as a point traveling on a trajectory provides a useful approximation for localized states, for intervals of time, and when entanglement can be neglected. A localized state may (mostly) pass through one slit of the interferometer but broadly supported states interact significantly with both slits. Quantum mechanics provides the general case while classical limits are more common in our experience. 

The Einstein-Podolsky-Rosen paradox [\ref{epr}], originally intended as a criticism of quantum mechanics, demonstrates that a classical concept for state, that is, an object with a determined, classically described state independent of the states of other objects, is not viable. A consequence of the observation that the principles of physics are independent of orientation is that angular momentum is conserved. If a spin zero particle decays into two spin one-half particles, then to conserve angular momentum the spins of the two product particles must add to zero. For any axis, one one-half spin particle must be spin up and the other spin down. Should these two particles fly apart, we can observe the spin of the near particle and infer the spin of the far distant particle. Should the near particle be spin up, then we know that to conserve angular momentum, the distant particle must be spin down. Or, if before the near measurement we rotate our measuring apparatus and measure spin of the near particle on another axis, it might be spin down or spin up on this new axis. But then the distant state must be in the corresponding spin up or down state on that axis. The issue is that if the description of the distant state has reality and is unaffected by our description of the near particle, how can angular momentum be conserved? How can our selection for measurement affect the distant description? The second particle is assumed to be sufficiently distant that the first particle does not causally affect the distant particle. If one accepts a classical description that the distant particle must be described by a determined spin, then conservation of angular momentum in quantum mechanics has a problem. Alternatively, the classical concept of a determined state for the distant particle in the pair has a problem. The quantum mechanical description is that the spin states of the pair are entangled as paired spins that conserve angular momentum. This entanglement was created when the original particle decayed and is not captured in a classical description. The observed states of the particle pair are described as correlated pairs, a description that preserves angular momentum. The classical idealization attributes reality to the spin of a distinguishable object independently of the states of other objects. This classical concept is not viable as a description of nature.

Another argument against classical descriptions of states is the celebrated Schr\"{o}dinger's cat paradox. To maintain the idea of a determined classical description and yet use the phenomenologically successful formalism of quantum mechanics necessitates ``collapse of the wave packet'' upon observation. This collapse from a quantum mechanical superposition of states to a classically described, determined state upon observation is imposed ad hoc to justify a classical description of the observer and the result of an observation. The Schr\"{o}dinger cat paradox takes consideration of a superposition of states from the microscopic, where it is less evident and more acceptable, to the macroscopic. Decay of an unstable isotope results in a dead cat in a box. There is a finite probability for the isotope to have decayed at any time. That is, at any time, the description of the quantum state includes components with decayed and undecayed isotopes. Schr\"{o}dinger's cat provides a second example of how the entanglement of states in quantum mechanics works. The live cat is entangled with the undecayed isotope, and the dead cat is entangled with the decayed isotope. An observation entangles states of the observer with one or the other correlated pair of states. Again, the classical concept that at any instance the cat is in a determined state, necessarily either alive or dead independently of the state of the observer, is not consistent with nature. Superposition applies and the quantum mechanical description includes entanglement of distinct states of the observer with the various states in the superposition of possibilities. This description is sometimes called a ``many-worlds'' interpretation, but bizarre images of splitting worlds does not apply. This relative state interpretation is completely consistent with our experience. The EWG, relative state interpretation, due to Hugh Everett, John Archibald Wheeler and Neill Graham [\ref{ewg}], demonstrates that there is no discernible difference between keeping a complete history of the evolution of the quantum state as a superposition of all possibilities from keeping only the history relative to one selected sequence of results of observations. The relevant physical description is completely indifferent to whether the alternative histories are accounted for or not. The alternative histories have no discernible effect on our future observations. This result follows from the Hilbert space description of states.

And finally, classical physics has its own flaws such as description of the electromagnetic radiation reaction force. These flaws degrade the plausibility of extrapolations of classical models to small scales.

\section{A construction of relativistic quantum physics}

The constructions provide alternative technical approaches to relativistic quantum physics [\ref{intro}]. The constructions result from consideration of possibilities within quantum mechanics that do not support the description of dynamics as canonical quantization. The constructions display causality, appropriate invariance under Poincar\'{e} transformations, nonnegative energy, and provide explicit Hilbert space realizations exhibiting interaction in physical spacetimes. The properties of these constructions suggest that difficulties in the union of relativity with quantum mechanics result from remnants of classical concepts in quantum mechanics and from an insufficient consideration for the properties of generalized functions.

Physical states are elements of a Hilbert space with a dense set of elements labeled by test function sequences. The Hilbert spaces appropriate for relativistic quantum physics derive from rigged (equipped) Hilbert spaces. A {\em rigged Hilbert space} is also designated a Gelfand triple after Israel Gelfand. In a Gelfand triple, a set of countably normed test functions are contained in a normed set of functions that label the elements of the Hilbert space, and the functions that label the elements of the Hilbert space are contained in the set of generalized functions (functionals) dual to the set of test functions. The Hilbert spaces result from isometries of equivalence classes of elements of the linear vector ({\em pre-Hilbert}) space of test function sequences to dense sets of elements for the Hilbert spaces. These sequences consist of functions with increasing numbers of spacetime arguments. This is entirely conventional. The revision is to select a subset of the generalized functions to consider and to label the states with functions that have Fourier transforms with support only on positive energies. That is, the Fourier transforms of the functions lack support on the negative energy support of the generalized functions that define a scalar product of functions sequences. The observation that nature includes only positive energies is satisfied by selecting the supports of the Fourier transforms of the functions that label physical states to be limited to positive energies. As a consequence, the support of these functions is not contained in a bounded region in spacetime. The scalar product provides a semi-norm on these sequences and implies the realization of states as elements of a Hilbert space. The implied scalar product of the Hilbert space is Poincar\'{e} invariant and local. Interaction is observed in the scalar products of plane wave limits of states. The selection of functions achieves concurrent satisfaction of the spectral support and microlocality conditions of relativistic quantum physics. The generalized functions that define the scalar product of the Hilbert space are the VEV of the fields and are generalized functions for an enveloping set of functions that includes the functions that lack negative energy support. The Fourier transforms of the enveloping set of functions are the span of products of multipliers of Schwartz functions of energy-momenta with Schwartz functions of momenta. The Fourier transforms of these functions are test functions in one less dimension than spacetime when $E^2=m^2+{\bf p}^2$ with $p:=E,{\bf p}$ the energy-momentum Lorentz vector. These functions include the spacetime Schwartz functions, generalized functions, for example, functions with temporal support concentrated at a point, and the functions used by Lehmann, Symanzik and Zimmermann in calculations of scattering amplitudes. The enveloping set of functions also includes functions of bounded spacetime support, used to define and test the local properties of the VEV. The quantum fields satisfy the established definition as multiplication in the algebra of function sequences but the algebra consisting of sequences of functions that lack negative energy support is not $*$-involutive. As a consequence, the real field does not satisfy Hermiticity, necessary to self-adjointness of a Hilbert space field operator, and it is unresolved whether the fields are generally Hilbert space operators. The ``operator-valued distributions'' $\Phi(x)$ are formally Hermitian, that is, real, but the lack of real functions among the state labels precludes Hermiticity of Hilbert space field operators. That the constructions do not satisfy a canonical quantization is emphasized by evaluation of the Hamiltonian. The Hamiltonian for the constructions with a single Lorentz scalar fields is $\sqrt{m^2+{\bf p}^2}$, assigned by canonical quantization as the Hamiltonian of a free field. As a consequence, the Lagrangian density associated with the construction is trivial while the development manifests non-trivial interaction. A description of quantum dynamics based upon quantization of classical Lagrangians excludes these constructions.


The constructions exhibit the physical properties of relativistic quantum physics but violate two prevailing technical assumptions. In the constructions:\begin{enumerate} \item the quantum fields are not self-adjoint operators \item no elements of the constructed Hilbert space are labeled by functions with support strictly limited to bounded regions of spacetime.\end{enumerate}The first technical difference results from the distinction between real fields and Hermitian Hilbert space field operators. The sets of functions that label the elements in the constructed Hilbert spaces lack real functions. The second difference is the result of the distinction between zero and arbitrarily small. Physically there is a negligible difference between zero and arbitrarily small but the implications of the distinction are decisive. The sets of functions that label states include dominantly-peaked packet states that are arbitrarily dominantly weighted within small bounded regions but never vanish in finite regions. There is an analogy with quasi-analytic functions: the entire function $\exp(-(z/\sigma)^2)$ is real for real values of $z$ when $\sigma \in {\bf R}$ and arbitrarily dominantly supported in a region of size proportional to $\sigma$, but the entire function does not vanish in any finite region. The only quasi-analytic function that vanishes in a finite region is zero. Acceptance of these two deviations from established technical conjecture results in realizations of interacting fields in physical spacetime. With these technical revisions, challenging `no go' theorems, in particular, demonstrations of the uniqueness of the two-point function ([\ref{feder}], the Jost-Schroer theorem and similar results [\ref{greenberg}]) do not apply to the constructions since the constructions lack Hermitian field operators, an assertion underlying the theorems. The constructions violate technical conjecture in the Wightman-G\aa rding, Wightman functional analytic, and Haag-Kastler algebraic axioms for relativistic quantum physics. The constructions violate aspects of each of the sets of axioms:\begin{itemize} \item[-] that fields are necessarily Hermitian Hilbert space operators in the Wightman-G\aa rding development. \item[-] that the spectral support condition and the semi-norm apply for all spacetime Schwartz functions in the Wightman functional analytic development. \item[-] that there are local observables strictly associated with bounded spacetime regions in the isotony condition of the Haag-Kastler development.\end{itemize}In the case of isotony, exclusion of the possibility of local observables has not been demonstrated but all the projections onto subspaces of states are not strictly localized. For the constructed examples of relativistic quantum physics, no states are labeled by functions of bounded spacetime support. The functions may be arbitrarily dominantly supported in a bounded spacetime region but the functions do not vanish in any region of spacetime, a property known as {\em anti-locality} [\ref{segal},\ref{masuda}]. Application of the spectral support condition and the semi-norm to spacetime Schwartz functions in the Wightman functional analytic development implies that real quantum fields are necessarily Hermitian Hilbert space operators. In contrast, the spectral support and the semi-norm axioms applied to the sets of functions with limited energy support do not imply that there are Hermitian Hilbert space field operators.

These two technical assertions, self-adjointness and bounded spacetime support, preclude the constructed realizations that exhibit interaction. The established axioms are too strong to admit the constructions. Indeed, the only realizations discovered for these axioms exhibit no interaction. The axioms exclude forms such as $\delta(p_1\!+\!p_2\!+\!\ldots p_n) \prod_{k=1}^n \delta(p_k^2-m^2)\,f(p_1,\ldots p_n)$ that are an evident choice for the Fourier transforms of non-trivial components of Lorentz scalar VEV and are generalized functions in four dimensions. $f$ is a symmetric, Lorentz invariant function. Nonnegative energy is established by the selection of functions that label the physical states. Such forms suffice physically and result in many Hilbert space realizations that exhibit the physical properties of relativistic quantum physics. The choice provided by the constructions is to either revise the axioms or to reject the sole elementary examples of relativistic quantum physics that exhibit interaction in physical spacetimes.

\section{Discussion}

Canonical quantization emphasizes extrapolation of a modeled classical limit back to specify the general case. Although extrapolation of classical methods both facilitated acceptance of the quantum theory and provides a method to predict and classify quantum dynamics, the method appears not to simply extend to include relativity and gravity. Constructions provide the alternative of quantum mechanical realizations with appropriate classical limits that are not canonical quantizations of classical models. A characterization of the explicit realizations of relativistic quantum physics that extend the constructions is an alternative to canonical quantization. The constructions apply selection of appropriate functions as state labels, functions in four spacetime dimensions with Fourier transforms supported only on positive energies, to realize relativistic quantum physics. These constructions, designated here as unconstrained QFT, are unconstrained by conjecture to implement canonical quantization.

An assumption that lingers in the established axioms for quantum mechanics is that observable, dynamic quantities correspond to self-adjoint Hilbert space operators. The constructions exploit that it is not necessary that classically observable quantities correspond to self-adjoint Hilbert space operators. More appropriate developments have been used in the practice of quantum mechanics but these refinements are not explicitly captured in axiomatic descriptions of quantum mechanics [\ref{dirac}].

Quantum mechanics includes the change from a description of nature as distinguishable geometric objects with determined positions in configuration spaces to a description of elements in Hilbert spaces. Quantum mechanics resolves observable flaws of classical mechanics, from an extensive entropy to conservation laws for quantized quantities to discrete atomic spectra. Despite this, classical descriptions continue to be considered foundations for quantum dynamics.

When canonical quantization is disregarded, the question of ``how is a classical interaction quantized'' is replaced by the more general ``what interactions are exhibited in the classical limits of quantum mechanics?'' This latter question results from consideration of explicit realizations of relativistic quantum physics such as the constructions. The question of whether there is a classical unification of the forces is not fundamental in this view of quantum mechanics without ``quantization''. In this view, no force is due to the geometry of spacetime and all forces are manifest in appropriate, classical limits of the quantum theory. Dynamics are determined given the vacuum expectation values (VEV) of the quantum fields. The VEV are generalized functions dual to the dense set of functions selected as state labels and VEV determine the scalar product in the Hilbert space. The intended analogy with aether questions the use of classical, geometric objects to describe relativistic quantum physics. Points or strings do not directly appear in a relativistic quantum mechanical development other than as interpretations of particular states. Difficulties in relativistic quantum physics may originate in the maintenance of familiar but ultimately unproductive concepts, analogous to aether. The risk from aether-like preconceptions is imposition of unnecessary constraints. The constructions are an example that realizes relativistic quantum physics by more fully exploiting quantum mechanical possibilities.

\begin{quote} Otherwise the ``paradox'' is only a conflict between reality and your feeling of what reality ``ought to be.'' Richard Feynman, sect.~18 p.~9 [\ref{feynman}].\end{quote}



\section*{References}
\begin{enumerate}
\item \label{gel2} I.M.~Gelfand, and G.E.~Shilov, {\em Generalized Functions, Vol.~2}, trans.~M.D.~Friedman, A.~Feinstein, and C.P.~Peltzer, New York, NY: Academic Press, 1968.
\item \label{wightman-hilbert} A.S.~Wightman,``Hilbert's Sixth Problem: Mathematical Treatment of the Axioms of Physics'', {\em Mathematical Development Arising from Hilbert Problems}, ed.~by F.~E.~Browder, {\em Symposia in Pure Mathematics 28}, Providence, RI: Amer.~Math.~Soc., 1976, p.~147.
\item \label{witten} {\em Quantum Fields and Strings: A Course for Mathematicians}, edited by P.~Deligne, D.~Kazhdan, P.~Etingof, J.W.~Morgan, D.S.~Freed, D.R.~Morrison, L.C. Jeffrey, E.~Witten, American Mathematical Society, 1999.
\item \label{weinberg} S.~Weinberg, {\em The Quantum Theory of Fields, Volume I, Foundations}, New York, NY: Cambridge University Press, 1995.
\item \label{intro} G.E.~Johnson, ``Introduction to quantum field theory exhibiting interaction'', Feb. 2015, arXiv:math-ph/\-1502.\-07727.
\item \label{limits} G.E.~Johnson, ``Classical limits of unconstrained QFT'', Dec. 2014, arXiv:math-ph/\-1412.\-7506.
\item \label{feymns} G.E.~Johnson, ``Fields and Quantum Mechanics'', Dec.~2013, arXiv:math-ph/\-1312.\-2608.
\item \label{wigner} T.D.~Newton and E.P.~Wigner, ``Localized States for Elementary Systems'', {\em Rev.~Modern Phys.}, Vol.~21, 1949, p.~400.
\item \label{yngvason-lqp} J.~Yngvason, ``Localization and Entanglement in Relativistic Quantum Physics'', Jan.~2014, arXiv:quant-ph/1401.2652.
\item \label{johnson} G.E.~Johnson, ``Measurement and self-adjoint operators'', May 2014, arXiv:quant-ph/\-1405.\-7224.
\item \label{ewg} B.S.~DeWitt, H.~Everett III, N.~Graham, J.A.~Wheeler in {\em The Many-worlds Interpretation of Quantum Mechanics}, ed.~B.S.~DeWitt, N.~Graham, Princeton, NJ: Princeton University Press, 1973.
\item \label{bogo} N.N.~Bogolubov, A.A.~Logunov, and I.T.~Todorov, {\em Introduction to Axiomatic Quantum Field Theory}, trans.~by Stephen Fulling and Ludmilla Popova, Reading, MA: W.A.~Benjamin, 1975.
\item \label{wizimirski} Z.~Wizimirski, ``On the Existence of a Field of Operators in the Axiomatic Quantum Field Theory'', {\em Bull. Acad. Polon. Sci., s\'{e}r. Math., Astr. et Phys.}, Vol.~14, 1966, pg.~91.
\item \label{borchers} H.J.~Borchers, ``On the structure of the algebra of field operators'', {\em Nuovo Cimento}, Vol.~24, 1962, p.~214.
\item \label{epr} A.~Einstein, B.~Podolsky, N.~Rosen, ``Can Quantum-Mechanical Description of Physical Reality be Considered Complete?'', {\em Phys.~Rev.}, Vol.~47, 1935, p.~777.
\item \label{feder} P.G.~Federbush and K.A.~Johnson, ``The Uniqueness of the Two-Point Function'', {\em Phys.~Rev.}, Vol.~120, 1960, p.~1926.
\item \label{greenberg} O.W.~Greenberg, ``Heisenberg Fields which vanish on Domains of Momentum Space'', {\em Journal of Math.~Phys.}, Vol.~3, 1962, pp.~859-866.
\item \label{segal} I.E.~Segal and R.W.~Goodman, ``Anti-locality of certain Lorentz-invariant operators'', {\em Journal of Mathematics and Mechanics}, Vol.~14, 1965, p.~629.
\item \label{masuda} K.~Masuda, ``Anti-Locality of the One-half Power of Elliptic Differential Operators'', {\em Publ. RIMS, Kyoto Univ.}, Vol.~8, 1972, p.~207.
\item \label{dirac} P.A.M.~Dirac, {\em The Principles of Quantum Mechanics}, Fourth Edition, Oxford: Clarendon Press, 1958.
\item \label{feynman} R.P.~Feynman, R.B. Leighton, and M.~Sands, {\em The Feynman Lectures on Physics, Volume III}, Reading, MA: Addison-Wesley Publishing Co., 1965.
\end{enumerate}
\end{document}